\documentstyle{article} 
\newcommand{\Ir}{Z\!\!\!Z}
\newcommand{\idty}{{\leavevmode{\rm 1\mkern -5.4mu I}}}
\newcommand{\Ibb}[1]{ {\rm I\ifmmode\mkern
            -3.6mu\else\kern -.2em\fi#1}}
\newcommand{\ibb}[1]{\leavevmode\hbox{\kern.3em\vrule
     height 1.2ex depth -.3ex width .2pt\kern-.3em\rm#1}}
\newcommand{\Cx}{{\ibb C}}
\newcommand{\Rl}{{\Ibb R}}
\newcommand{\Nl}{{\Ibb N}}
\newcommand{\be}{\begin{equation}}
\newcommand{\ee}{\end{equation}}
\newcommand{\bea}{\begin{eqnarray}}
\newcommand{\eea}{\end{eqnarray}}
\newcommand{\A}{{\cal A}}
\renewcommand{\H}{{\cal H}}
\renewcommand{\O}{\Omega}

\newcommand{\hpa}{\hat{\partial}}
\renewcommand{\d}{{\rm d}}
\renewcommand{\a}{{\bf a}}

\begin{document}

\begin{center}
\large \bf NONCOMMUTATIVE GEOMETRY AND A CLASS 
       OF COMPLETELY INTEGRABLE MODELS 
\end{center}
\vskip.2cm
\begin{center} {\bf A. Dimakis} 
\vskip.1cm
 Department of Mathematics, University of the Aegean \\
               GR-83200 Karlovasi, Samos, Greece 
\vskip.1cm
                       and 
\vskip.1cm
         {\bf F. M\"uller-Hoissen} 
\vskip.1cm
Max-Planck-Institut f\"ur Str\"omungsforschung \\
      Bunsenstrasse 10, D-37073 G\"ottingen, Germany
\end{center}

\vskip.2cm

\begin{abstract}
We introduce a Hodge operator in a framework of noncommutative geometry.
The complete integrability of 2-dimensional classical harmonic maps 
into groups ($\sigma$-models or principal chiral models) is then extended 
to a class of  `noncommutative' harmonic maps into matrix algebras. 
\end{abstract}

\section{Introduction}
A generalization of classical (pseudo-) Riemannian geometries is obtained 
by generalizing the concept of differential forms, accompanied with a 
suitable generalization of the Hodge operator.
The algebra of (ordinary) differential forms is replaced by a 
differential algebra on some, in general noncommutative, algebra. In this
setting one can consider `noncommutative' analogues of physical models 
and dynamical systems. After collecting some basic
definitions in section 2, section 3 presents a class of completely
integrable generalized harmonic maps into matrix algebras which are
noncommutative analogues of harmonic maps into groups (also known as 
$\sigma$-models or principal chiral models). 
This extends our previous work [1-4] where, in particular, the nonlinear
Toda lattice has been recovered as a generalized harmonic map with respect
to a `noncommutative differential calculus' on $\Rl \times \Ir$. 
Section 4 contains some conclusions.

\section{Basic definitions}

\subsection{Differential calculus over associative algebras}
Let $\A$ be an associative algebra over $\Cx$ (or $\Rl$) with unit element $\idty$. 
A {\em differential algebra} is a $\Ir$-graded associative algebra (over $\Cx$, 
respectively $\Rl$) $ \O (\A) = \bigoplus_{r \geq 0} \O^r (\A)$ 
where the spaces $\O^r (\A)$ are $\A$-bimodules and $\O^0(\A) = \A$. 
A {\em differential calculus} over $\A$ consists of a differential algebra $\O(\A)$ 
and a linear\footnote{Here {\em linear} means linear over $\Cx$, respectively $\Rl$.} 
map $ \d \, : \,  \O^r (\A) \rightarrow \O^{r+1}(\A)$ with the properties
\bea
       \d^2 = 0 \, , \qquad
       \d (w \, w') = (\d w) \, w' + (-1)^r \, w \, \d w'                 
\eea
where $w \in \O^r(\A)$ and $w' \in \O (\A)$. The last relation is
known as the (generalized) {\em Leibniz rule}. We also require 
$\idty \, w = w \, \idty = w$ for all elements $w \in \O (\A)$. The 
identity $\idty \idty = \idty$ then implies $ \d \idty = 0 $. Furthermore,
it is assumed that $\d$ generates the spaces $\O^r(\A)$ 
for $r>0$ in the sense that $\O^r(\A) = \A \, \d \O^{r-1}(\A) \, \A$.

\subsection{Hodge operators on noncommutative algebras}
Let $\A$ be an associative algebra with unit $\idty$ and an  involution ${}^\dagger$. 
Let $\O(\A)$ be a differential calculus over $\A$ such that there exists 
an invertible map $ \star \, : \,  \O^r(\A) \rightarrow \O^{n-r}(\A) $
for some $n \in \Nl$, $r = 0, \ldots, n$, with the property
\be                 \label{star_cov_dag}
  \star \, (w \, f) = f^\dagger \, \star w    \qquad  \forall w \in \O^r(\A), \, f \in \A  \; .
 \ee
Such a map $\star$ is called a (generalized) {\em Hodge operator}.\footnote{As 
a consequence, the inner product $\O^1(\A) \times \O^1(\A) \rightarrow \Cx$
defined by $ (\alpha , \beta) = \star^{-1} (\alpha \star \beta)$ satisfies
$ (\alpha , \beta \, f) = (\alpha \, f^\dagger, \beta)$ and $ (f \, \alpha , \beta) = 
(\alpha , \beta) \, f^\dagger$.}
We will furthermore assume that ${}^\dagger$ extends to an involution of $\O(\A)$ 
so that
\be
                 (w \, w')^\dagger = {w'}^\dagger \, w^\dagger   \; .
\ee
Then the further condition
\be                             
     (\star \, w)^\dagger = \star^{-1}(w^\dagger)         \label{star_dag}
\ee
can be consistently imposed on the calculus, since 
\be
        ( \star \, (w \, f) )^\dagger = (f^\dagger \, \star w)^\dagger 
                                           = (\star \, w)^\dagger \, f
                                           = [\star^{-1}(w^\dagger)] \, f
                                           = \star^{-1}(f^\dagger w^\dagger)
                                           = \star^{-1}[(w f)^\dagger]  \; .
\ee
We still have to define how the exterior derivative $\d$ interacts with the involution.
Here we adopt the rule\footnote{See also [5]. A different  though equivalent extension
of an involution on $\A$ to $\O(\A)$ was chosen in [6]: $(w \, w')^\ast = 
(-1)^{rs} w'^\ast \, w^\ast$ where $w \in \O^r(\A)$, $w' \in \O^s(\A)$, and 
$(\d w)^\ast = \d (w^\ast)$. The two extensions are related by 
$w^\ast = (-1)^{r(r+1)/2} w^\dagger$.} 
\be        \label{d_dag}
      (\d w)^\dagger = (-1)^{r+1} \, \d (w^\dagger)       \qquad   w \in \O^r(\A) \; .
\ee

\subsection{Noncommutative harmonic maps into matrix algebras}
Let $\A$ be an associative algebra with unit $\idty$ and $\H$ an algebra 
generated by the entries $\a^i{}_j \in \A$, $i, j = 1 \ldots, N$, of a 
matrix $\a$ with generalized inverse\footnote{Examples are given by matrix
Hopf algebras (cf [6]) in which case the antipode provides us with a generalized inverse.} 
$S$, i.e., 
\be
   S(\a^i{}_k) \, \a^k{}_j
      = \delta^i_j \, \idty 
      = \a^i{}_k \, S(\a^k{}_j)   \; .
\ee
Given a differential calculus $(\O(\A), \d)$, the matrix of 1-forms
\be        \label{A_pure_gauge}
               A := S(\a) \, \d \a
\ee 
satisfies the (zero curvature) identity
\be      \label{F=0}
          F := \d A + A A = 0  \; .
\ee
Let us now assume that $(\O(\A), \d)$ admits a Hodge operator $\star$.
The equation
\be
       \d \star A = 0      \label{hm_eq}
\ee
then defines a {\em generalized harmonic map} into a matrix algebra.\footnote{We may 
also call this a {\em generalized principal chiral model} or a {\em generalized $\sigma$-model}. 
Actually, we only need the restriction of the Hodge operator to 1-forms here, i.e.,  
$\star \, : \; \O^1(\A) \rightarrow \O^{n-1}(\A)$.}
\vskip.2cm

A {\em conserved current} of a generalized harmonic map is a 1-form $J$ which satisfies 
$ \d \star J = 0 $ as a consequence of (\ref{hm_eq}). 
We call a generalized harmonic map {\em completely integrable} if there is an infinite
set of independent\footnote{A convenient notion of independence in this context still has 
to be found.} conserved currents.

\section{Completely integrable 2-dimensional generalized harmonic maps}
 For 2-dimensional classical $\sigma$-models there is a construction of an
infinite tower of conserved currents [7]. This has been generalized
in [1-4] to harmonic maps on ordinary (topological) spaces, 
but with {\em non}commutative differential calculi, and values in a matrix group. 
In the following, we present another generalization to harmonic maps
on, in general {\em non}commutative algebras (see also [4]). 
\vskip.1cm

Let $(\O(\A), \d)$ be a differential calculus over an associative algebra $\A$ with 
unit $\idty$, involution ${}^\dagger$ and a Hodge operator $\star$ satisfying 
the rules listed in section 2.2 with $n=2$. Furthermore, let us consider a generalized
harmonic map into a matrix algebra. If in addition the following conditions are 
satisfied, then the construction of an infinite tower of conservation laws (for classical
$\sigma$-models) mentioned above also works in the generalized setting
under consideration.\footnote{For {\em commutative} algebras, less restrictive
conditions were given in [1,2].}
\begin{enumerate}
\item For each $r = 0,1,2$ there is a constant $\epsilon_r \neq 0$ such that
\be     \label{star_eps}
   \star \star w = \epsilon_r \, w  \qquad \forall w \in \O^r \; .
\ee
Using $\star \star (\star \, w) = \star \, (\star \star \, w)$ we find
\be
       \epsilon_{2-r} = \epsilon_r^\dagger  \; .
\ee
(\ref{star_eps}) together with (\ref{star_dag}) in the form 
$\star \star (\star \star \, w)^\dagger = w^\dagger$ leads to
\be    \label{eps_dag}
        \epsilon_r^\dagger =  \epsilon_r^{-1}    \; .
\ee
In particular, it follows that $\epsilon_2 = \epsilon_0^\dagger = \epsilon_0^{-1}$
and $\epsilon_1 = \pm 1$.
\item We impose the modified symmetry condition\footnote{As a consequence, 
the inner product defined in a previous footnote satisfies $ (\alpha , \beta)^\dagger  = 
(\beta , \alpha)$.}  
\be                  \label{dag_star_eps}
     (\alpha \star \beta)^\dagger = \epsilon_0 \; \beta \star \alpha                                                
\ee
where $\alpha, \beta \in \O^1(\A)$. This is consistent with (\ref{star_cov_dag}) 
since
\be
      [ \alpha \star (\beta \, f) ]^\dagger
   = [ \alpha \, f^\dagger \, \star \beta]^\dagger 
   = \epsilon_0 \, \beta \star (\alpha \, f^\dagger)
   = \epsilon_0 \, (\beta \, f) \star \alpha          \; .
\ee 
\item $\epsilon_0 = - \epsilon_1$.
\item The first cohomology is trivial, i.e., for $\alpha \in \O^1(\A)$ we have
\be              \label{closed-star-star} 
   \d \alpha = 0 \quad \Rightarrow \quad  \exists \, \chi \in \A \, : \quad
       \alpha = \d \chi  \; .
\ee
\end{enumerate}
As a consequence of (\ref{hm_eq}),
\be
    J^{(1)} := D \chi^{(0)} = (\d + A) \, \chi^{(0)} = A
  \quad \mbox{where} \quad   \chi^{(0)} := \mbox{diag}(\idty, \ldots, \idty)
\ee
is conserved. Let $J^{(m)}$ be any conserved current.
Using (\ref{star_eps}) and (\ref{closed-star-star}), this implies
\be        \label{J_star_chi}
     J^{(m)} = \star \, \d (\chi^{(m)\dagger})
\ee 
with an $N\times N$ matrix $\chi^{(m)}$. Now
\be       \label{J_Dchi}
     J^{(m+1)} := D \chi^{(m)}  
\ee
is also conserved, since
\bea
     \d \star J^{(m+1)} 
   &=& \d \star D \chi^{(m)} 
   = -\epsilon_1 \, [D \star \d (\chi^{(1)\dagger}]^\dagger	
   = -\epsilon_1 \, [D J^{(m)}]^\dagger  \nonumber \\
   &=& -\epsilon_1 \, [DD \chi^{(m-1)}]^\dagger
   = -\epsilon_1 \, [F \chi^{(m-1)}]^\dagger
   = 0 \; .
\eea
The second equality in the last equation follows from the next result.
\vskip.2cm
\noindent
{\em Lemma.} For a matrix $\chi$ with entries in $\A$ we have
\be
 \d \star D \chi = - \epsilon_1 \, (D \star \d (\chi^\dagger))^\dagger   \; .
\ee
{\em Proof:} First we note that
\begin{eqnarray*}
       ( \d \star \d \chi^i{}_j )^\dagger 
  =   \d (\star \, \d \chi^i{}_j)^\dagger 
  =   \d \star^{-1} (\d \chi^i{}_j)^\dagger
  =  - \d \star^{-1} \d (\chi^i{}_j)^\dagger  
  = - \epsilon_1 \, \d \star \d (\chi^i{}_j)^\dagger
\end{eqnarray*}
using (\ref{d_dag}), (\ref{star_dag}), again (\ref{d_dag}), then 
(\ref{star_eps}) and (\ref{eps_dag}). Furthermore,
\begin{eqnarray*}
       [ \d (\chi^k{}_j)^\dagger \star A^i{}_k ]^\dagger  
    = \epsilon_0 \, A^i{}_k \star \d (\chi^k{}_j)^\dagger  
\end{eqnarray*} 
using (\ref{dag_star_eps}). Hence
\begin{eqnarray*}
    \d \star D \chi^i{}_j & = & \d \star (\d \chi^i{}_j + A^i{}_k \, \chi^k{}_j) 
    = \d \star \d \chi^i{}_j + \d ((\chi^k{}_j)^\dagger \star A^i{}_k)   \nonumber \\
 & = & \d \star \d \chi^i{}_j + \d (\chi^k{}_j)^\dagger \star A^i{}_k +
           (\chi^k{}_j)^\dagger \d \star A^i{}_k   \nonumber  \\
 & = & [ (\d \star \d \chi^i{}_j)^\dagger 
        + (\d (\chi^k{}_j)^\dagger \star A^i{}_k)^\dagger ]^\dagger 
                                                                 \nonumber \\
 & = & [- \epsilon_1 \,	\d \star \d(\chi^i{}_j)^\dagger 
        + \epsilon_0 \, A^i{}_k \star \d (\chi^k{}_j)^\dagger ]^\dagger
\end{eqnarray*}
using $\d \star A = 0$. Inserting $ \epsilon_0 = - \epsilon_1$ now 
completes the proof.
\hfill  {\Large $\Box$}
\vskip.2cm
\noindent
In this way we obtain an infinite set of (matrices of) conserved currents. 
Introducing $\chi := \sum_m \lambda^m \, \chi^{(m)}$ with a parameter 
$\lambda \in \Cx$, (\ref{J_star_chi}) together with (\ref{J_Dchi}) leads to the 
linear equation
\be           \label{master_eq}
      \star \, \d (\chi^\dagger) = \lambda \, D \chi    
\ee
(see also [2]). As a consequence, $D \star \d \chi^\dagger 
= \lambda \, D^2 \chi = \lambda \, F \chi$ and
$0 = (\d \star \, D \chi)^\dagger = - \epsilon_1 \, D \star \d \chi^\dagger
     + (\d \star A)^\dagger \, \chi $,
from which the following integrability condition is obtained,
\be
    [ (\d \star A)^\dagger - \epsilon_1 \, \lambda \, F ] \, \chi = 0 \; .
\ee
If $F = 0$, which is solved by (\ref{A_pure_gauge}), then
the harmonic map equation $\d \star A = 0$ results. Alternatively, 
$\d \star A = 0$ is solved by $A = \star \, \d (\phi^\dagger)$ with a matrix 
$\phi$ with entries in $\A$. Then the integrability condition becomes
\be
   0 = F = \d \star \, \d \phi^\dagger 
      + (\star \, \d \phi^\dagger) (\star \, \d \phi^\dagger)
      = \d \star \, \d \phi^\dagger - \d \phi \, \d \phi
\ee
using (\ref{dag_star_eps}), (\ref{star_eps}) and $\epsilon_0 \, \epsilon_1 = -1$.

\subsection{Examples}
\noindent
(1) Let $\A$ be the Heisenberg algebra with the two generators 
$q$ and $p$ satisfying $[q,p] = i \, \hbar$.
In the simplest differential calculus over $\A$ we have 
$[\d q , f] =0$ and $[\d p , f] = 0$ for all $f \in \A$. 
It follows that $ \d f  = (\hpa_q f) \, \d q + (\hpa_p f) \, \d p$
where the generalized partial derivatives are given by
\be
         \hpa_q f := -{1\over i\hbar}[p,f] \, ,   \qquad
         \hpa_p f := {1\over i \hbar}[q,f]   \; .
\ee
Acting with $\d$ on the above commutation relations for `functions' and
differentials, one obtains $ \d q \, \d q = 0$, $\d q \, \d p + \d p \, \d q =0$ 
and $ \d p \, \d p = 0 $. As an involution we choose hermitean conjugation with
$q^\dagger = q, \, p^\dagger = p$. A Hodge operator satisfying the conditions 
(\ref{star_dag}), (\ref{star_eps}) and (\ref{dag_star_eps}) is determined by
\be
   \star \, 1 = \d q \, \d p \, , \quad 
   \star \, \d q = \d p  \, ,  \quad 
   \star \, \d p = \d q \, , \quad
   \star \, (\d q \, \d p) = - 1   \, ,
\ee
so that $\epsilon_0 = \epsilon_2 =-1$ and  $\epsilon_1 = 1$. 
Now we consider a generalized harmonic map with values in the group 
of unitary elements $U$ of $\A$ 
which satisfy $U^\dagger U= \idty =U U^\dagger$. With
\be
      A = U^\dagger \, \d U 
        = - {1\over i \hbar} \, (U^\dagger p \, U-p) \, \d q 
          + {1 \over i \hbar} \, (U^\dagger q \, U-q) \, \d p
\ee
we get $ \star \, A =  (i \hbar)^{-1} \, (U^\dagger p \, U-p) \, \d p 
 - ( i \hbar)^{-1} \, (U^\dagger q \, U-q) \, \d q $ and the harmonic map 
equation $\d \star A = 0$ becomes
$ [p , U^\dagger p \, U] - [q , U^\dagger q \, U] = 0 $.
In terms of $P := U^\dagger p \, U$ and $Q := U^\dagger q \, U$ this 
takes the form
\be  
  [p,P] - [q,Q] = - i \, \hbar \, (\hpa_q P + \hpa_p Q) = 0 \; .  
\ee
On the level of formal power series in $q$ and $p$, every closed 1-form 
is exact so that (\ref{closed-star-star}) holds. All required conditions are 
fulfilled in this example. From (\ref{master_eq}) one derives
\be
  \d \chi = \lambda \, (1- \lambda^2)^{-1} \, (\lambda \, A - \star A) \, \chi  
\ee
using $A^\dagger = A$. Reading off components with respect to the basis 
$\lbrace \d q, \d p \rbrace$ of $\O^1(\A)$ leads to $ \chi \, q = L \, \chi $ 
and $ \chi \, p = M \, \chi $ where
\be
    L := {\lambda \over 1- \lambda^2} \, (\lambda^{-1} \, q - p 
           - \lambda \, Q + P), \quad 
    M :=  {\lambda \over 1- \lambda^2} \, (-q + \lambda^{-1} \, p 
             + Q - \lambda \, P) \, .
\ee
The integrability condition is then $[L,M] = i \hbar$.
\vskip.2cm
\noindent
(2) Let $\A = C^\infty(\Rl^2)$ with the (noncommutative) Moyal product [8] 
\be
     f \ast h = m \circ e^{(i \hbar/2) \, P} \, (f \otimes h)   
\ee
where $P := \partial_q \otimes \partial_p - \partial_p \otimes \partial_q$
in terms of real coordinates $q$ and $p$, and $m(f \otimes h) = f h$ for
$f,h \in \A$. An involution is given by
$ (f \ast h)^\dagger = h^\dagger \ast f^\dagger $
where ${}^\dagger$ acts as complex conjugation on the functions. In terms of
the real generators $q$ and $p$ of $\A$, the simplest differential 
calculus\footnote{Differential calculi on the Moyal algebra were also considered
in [9].}
is determined by $\d q \ast f = f \ast \d q$ and $\d p \ast f = f \ast \d p$
as in our first example. A Hodge operator is then given by
\be
   \star \, 1 = \d q \ast \d p \, , \quad 
   \star \, \d q = \d p  \, ,  \quad 
   \star \, \d p = \d q \, , \quad
   \star \, (\d q \ast \d p) = - 1  \; .
\ee
The differential calculus has trivial cohomology and all required conditions are
satisfied. (\ref{hm_eq}) implies $A = \star \, \d (\phi^\dagger) $ with a matrix
$\phi$ with entries in $\A$. The harmonic map equation is then obtained
by substituting this expression into the zero curvature condition (\ref{F=0}).
We obtain
\be
  (\partial_q^2 - \partial_p^2) \phi - m \circ e^{(i \hbar/2) \, P} 
  \, P(\phi \otimes \phi)  = 0
\ee  
which is
\be
   \Box \phi - \partial_q \phi \ast \partial_p \phi 
   + \partial_p \phi \ast \partial_q \phi =0   \; .
\ee
This is a deformation of a classical principal chiral model. For $\phi \in \A$
it is a deformation of the wave equation.

\section{Conclusions}
We introduced Hodge operators and a class of harmonic maps on
(suitable) differential calculi on associative algebras. 
 Furthermore, we generalized a construction of an infinite tower of 
conserved currents from the classical framework of $\sigma$-models
[7] to this framework of noncommutative geometry.  
It involves a drastic generalization of a notion of `complete integrability'. 
This is a very peculiar property of an equation and we have presented
a constructive method to determine corresponding equations. Further
elaboration of examples is necessary to clarify their significance, however.

\vskip.3cm
\noindent
 F. M.-H. is grateful to C. Burdik for the kind invitation to present 
the material of this paper at the $7^{th}$ Colloquium on {\it Quantum Groups 
and Integrable Systems}.

\end{document}